\documentclass[]{JHEP3} 
\usepackage[dvips]{graphicx}
\usepackage{datetime}

\usepackage{epsfig,multicol,bbm}
\usepackage{amssymb}
 \def\dst{\displaystyle}

\title{Predictions for TOTEM experiment at LHC  from integral and derivative dispersion relations}

\author{A. Alkin and E. Martynov \\
Bogolyubov Institute for Theoretical Physics, National Academy of Sciences of Ukraine, Metrologichna 14b, Kiev, Ukraine, 03680\\
	E-mail: \email{alkin@bitp.kiev.ua}\\
E-mail: \email{martynov@bitp.kiev.ua}}

\abstract{
Integral and derivative dispersion relations (IDR and DDR) are
considered for the forward scattering $pp$ and $\bar pp$ amplitudes. A
scheme for calculation of the corrections to asymptotic form for the DDR is presented.
The data on the total cross sections for $pp (\bar pp)$ interaction as well as the data on the ratios
$\rho $ of real to imaginary part of $pp$ and $\bar pp$ forward  scattering amplitudes have been analyzed
by the IDR and DDR methods within high-energy Regge models.
Regge parametrizations of high-energy total cross sections supplemented by the properly calculated low-energy part of the
dispersion integral (from the two-proton threshold up to $\sqrt{s}=$5 GeV)
allow to reproduce well the $\rho $ data at low energies with
the only free parameter, subtraction constant. It is
shown that three models for the pomeron, simple pole pomeron (with intercept $\alpha_{P}(0)>1$), tripole
pomeron and dipole pomeron (the both with $\alpha_{P}(0)=1$) lead
to practically equivalent description of the available data at $\sqrt{s}>$5 GeV. Predictions for the TOTEM measurements of $\sigma_{t}$ and $\rho$ at $\sqrt{s}=$ 7 and 14 GeV are given.
}

\begin{document}


\section{Introduction}
The energy dependence of the hadronic total cross sections as well as
of the parameters $\rho=Re A(s,0)/Im A(s,0)$ - the ratios of the real to
the imaginary part of the forward scattering amplitudes - was widely discussed and has a long history. The problem was considered on the base of integral dispersion relations (IDR) \cite{GNU, BlockCahn, Soding, Amaldi, UA4/2, CMS-1, CMS-2, AvilaMenon}, derivative dispersion relations (DDR) \cite{CMS-1, CMS-2, GribovMigdal, BronzKaneSukh,  KangNicol} as well as within the various analytical high-energy models \cite{DGM, COMPETE, COMPETE-1,COMPETE-2, d-l}.

However, in spite of a big activity in this area situation remains somewhat undecided.
In the papers of COMPETE collaboration \cite{COMPETE, COMPETE-1,COMPETE-2}, all available data on $\sigma_{tot}(s)$ and
$\rho(s)$ for hadron-hadron, photon-hadron and photon-photon interactions were considered. Many analytical models for the forward scattering
amplitudes were fitted and compared. The ratio $\rho$ was calculated in explicit form, from the amplitudes parameterized explicitly
by contributions from the pomeron and secondary reggeons. The values of the free parameters were determined from the fit to the data at
$s\ge s_{min}$, where $\sqrt{s_{min}}=5$ GeV. Omitting all details, we note here the main two conclusions.
The best description of the data (with the minimal $\chi^{2}/dof$, where $dof$ is the number of degrees of freedom) is obtained
for the model with $\sigma_{tot}$ rising as $\log^{2}s$ at $s\to \infty$. The dipole pomeron model ($\sigma_{tot}(s)\propto \ln s$)
also lead to good description while the simple pomeron model with $\sigma_{tot}(s)\propto s^{\epsilon}, \epsilon >0$   was excluded from the
list of the best models (in accordance with COMPETE criteria, see details in \cite{COMPETE-1}).

Analysis of these results shows that the simple pomeron model was rejected due to a poor
description of $\rho$ data at low energy. On the other hand, there are a few questions concerning the explicit Regge-type
models usually used for analysis and description of the data.

How low in energy can the Regge parameterizations be extended, as they are written as functions of the asymptotic variable $s$ rather than
the "Regge" variable $z_{t}=\cos\theta_{t}$ ($|z_{t}|=E/m$ in the laboratory system for identical colliding particles)?

At which energies can the "asymptotic" normalization
\begin{equation}\label{eq:opt-t-as}
\sigma_{tot}(s)=\frac{1}{s}\Im mA(s,0)
\end{equation}
be used instead of the standard one?
\begin{equation}\label{eq:opt-t-st}
\sigma_{tot}(s)=\frac{1}{2mp_{l}}\Im mA(s,0)
\end{equation}
 In the above equation $m$ is a mass of the target hadron and $p_{l}$ is a momentum of the  incoming hadron at laboratory system.

And last, how do the value of the ratio $\rho$ obtained from the derivative dispersion relation deviates from those calculated
through the dispersion integral?

In this paper, we try to answer these questions considering the above mentioned three pomeron models for $pp$ and $\bar pp$
interactions at $\sqrt{s}\geq 5$ GeV within the dispersion relation methods. We compare the methods of integral and derivative
dispersion relations and of  model parametrization giving explicitly the both real and imaginary part of scattering amplitude. We give as well the predictions for
$\sigma_{tot}(s)$ and $\rho(s)$ at the LHC energies of the pomeron models obtained by the integral dispersion relation method.

\section{Integral and derivative dispersion relations.}
Assuming, in accordance with many analyzes, that the odderon does not
contribute asymptotically at $t=0$, one can show that the integral
dispersion relations (IDR) for $pp$ and $\bar pp$ amplitudes can be
reduced to those with one subtraction constant \cite{Soding}:
\begin{equation}\label{eq:dispers II}
  \rho_{\pm}\sigma_{\pm}=\dst \frac{B}{p} +
  \displaystyle \frac{E}{\pi p}\,{\rm P}\!\int\limits_{m_{p}}^{\infty}\left
[\frac{\sigma_{\pm}}{E'(E'-E)}-\frac{\sigma_{\mp}}{E'(E'+E)}\right ]p'\,
dE'
\end{equation}
where $m_{p}$ is the proton mass, $E$ and $p$ are the energy and momentum
of the proton in the laboratory system, and $B$ is a subtraction constant,
usually determined from the fit to the data. The indices $+(-)$ stand
respectively for the $\bar pp$ and $pp$ amplitudes. The standard
normalization (\ref{eq:opt-t-st}) is chosen in Eq.(\ref{eq:dispers
II}).

In the above expression, the contributions of the integral over the nonphysical cuts from the two-pion to the two-proton threshold are omitted
because they are $\lesssim 1\%$ (see, e.g. \cite{valeng}) in the region of interest ($\sqrt{s}\ge 5$ GeV).

For the first time the dispersion relations (\ref{eq:dispers II}) were used by P. S\"{o}ding \cite{Soding} to analyze low energy data at $\sqrt{s}<$ 6 GeV with a
numerical integration of the $\sigma$ data at $\sqrt{s}<$ 4.7 GeV.  Then U. Amaldi et. al. \cite{Amaldi} and UA4/2 Collaboration \cite{UA4/2}  applied  IDR  to $pp$
and $\bar pp$ data when the cross section at ISR and SPS energies were measured. Calculating the dispersion integral they used a high energy model for cross
section just from the threshold $E=m_{p}$. Strictly speaking it could not be correct. Nevertheless a reasonable description of high energy data was obtained.
Probably it took place  due to fitted substraction constant which somehow compensates  an incorrect contribution of low energy part of the dispersion integral. In
the Section \ref{sec:low} we  apply another, more accurate, suggested in \cite{CMS-1, CMS-2} way for calculation of low energy part of dispersion integral.

An alternative way is to consider so called derivative dispersion relations (DDR) \cite{GribovMigdal, BronzKaneSukh} which were obtained
separately for crossing-even and crossing-odd amplitudes
\begin{equation}\label{eq:crossampl}
f_{\pm}(s,0)=A_{+}(s,0)\pm A_{-}(s,0)
\end{equation}
which have the following crossing symmetry properties
\begin{equation}\label{eq:crossprop}
f_{\pm}(s,0)=\pm f_{\pm}(4m_{p}^{2}-s,0).
\end{equation}
In particular, in \cite{BronzKaneSukh} DDRs were written in the following asymptotic (at $E/m_{p}\to \infty$) forms
{\small
\begin{equation}\label{eq:ddr-bks}
\begin{array}{ll}
\Re ef_{+}(E,0)=&\dst (E/m_{p})^{\alpha}\tan \left [\frac{\pi}{2}\left (\alpha-1+E\frac{d}{dE}\right )\right ] \frac{\Im mf_{+}(E,0)}{(E/m_{p})^{\alpha}} \\
\Re ef_{-}(E,0)=&\dst (E/m_{p})^{\alpha}\tan \left [\frac{\pi}{2}\left (\alpha+E\frac{d}{dE}\right )\right ] \frac{\Im mf_{-}(E,0)}{(E/m_{p})^{\alpha}} \end{array}
\end{equation}}
There were attempts to relate the parameter $\alpha $ in Eq.~(\ref{eq:ddr-bks}) with a reggeon intercept. However, it can be easily proved that the right-hand
parts of the equations in (\ref{eq:ddr-bks}) do not depend on $\alpha$.

Indeed, let us consider the function
$$
G(x,\nu)=\Phi(\nu+d/dx)\phi (x)e^{-\nu x}
$$
If function  $\Phi(z)$ can be expanded in the power series of $z$ then
\begin{equation}\label{eq:G(x,nu)}
\begin{array}{ll}
G(x,\nu)&=\sum\limits_{k=0}^{\infty}\frac{\Phi^{(k)}(0)}{k!}\left
(\nu+\frac{d}{dx}\right )^{k}\phi (x)e^{-\nu x}
= \sum\limits_{k=0}^{\infty}\frac{\Phi^{(k)}(0)}{k!}\left
(\nu+\frac{d}{dx}\right )^{k-1}\left
(\nu+\frac{d}{dx}\right )\phi (x)e^{-\nu x}\\
&=\sum\limits_{k=0}^{\infty}\frac{\Phi^{(k)}(0)}{k!}\left
(\nu+\frac{d}{dx}\right )^{k-1}e^{-\nu x}\left [
\frac{d}{dx}\right ] \phi (x)=\ldots \\
&=e^{-\nu x}\sum\limits_{k=0}^{\infty}\frac{\Phi^{(k)}(0)}{k!}\left [
\frac{d}{dx}\right ]^{k}\phi (x)=e^{-\nu x}\Phi\left (\frac{d}{dx}\right )\phi (x).
\end{array}
\end{equation}
Making use of this property of $G(x,\nu)$ one can see that the Eqs.~(\ref{eq:ddr-bks}) are valid at any value of $\alpha$. Consequently they can be rewritten (putting
$\alpha=1$ and $\alpha=0$ in expressions for $f_{+}$ and $f_{-}$ correspondingly) in the form
\begin{equation}\label{eq:ddrev-as}
\begin{array}{ll}
\Re ef_{+}(E,0)\approx &E\tan \left [ \frac{\pi}{2}E\frac{d}{dE}\right ]\Im mf_{+}(E,0)/E,\\
\Re ef_{-}(E,0)\approx &\tan \left [ \frac{\pi}{2}E\frac{d}{dE}\right ]\Im mf_{-}(E,0).
\end{array}
\end{equation}

DDR are very useful in practice due to their simple analytical form at high energies, $E\gg m_{p}$. However it would be of importance to estimate the corrections to these asymptotic relations (\ref{eq:ddrev-as}) while they are applied at finite $s$. The method to find all corrections has been developed  in  the \cite{CMS-1, CMS-2}.  The dispersion integral can be transformed into the series (subtraction constant is omitted for the moment as well  the additional term $\frac{E}{\pi m_{p}} \ln\left
(\frac{E+m_{p}}{E-m_{p}}\right )\Im mf_{+}(m_{p},0)$ which is supposed to be equal to zero if the imaginary part of amplitude vanishes at the threshold):
$$
\Re ef_{+}(E,0)=
 \frac{2E}{\pi m_{p}}\sum\limits_{p=0}^{\infty}\frac{1}{2p+1}
\sum\limits_{k=0}^{\infty}{\cal I}(\xi;p,k)\frac{ g^{(k+1)}(x)}{k!}\bigg |_{x=\xi}
$$
where $\xi=\ln(E/m_{p})$,  $g^{(k)}(x)$ is the $k$-th derivative of the function $g(x)$,
$$g(\xi)= m_{p}\Im
mf_{+}(E,0)/E,
$$
$$
{\cal I}(\xi;p,k) 
={(2p+1)^{-k-1}}\left [\Gamma (k+1)+ (-1)^{k}(2p+1)\gamma(k+1,\xi)
\right ] 
$$
and $\gamma(a,x)$ is an incomplete gamma function.

The first term gives exactly the asymptotic expression (\ref{eq:ddrev-as}). The second term after some transformations leads to the series of corrections presented
in the complete expression for real part of the crossing-even amplitude
\begin{equation}\label{eq:ddreven}
\Re ef_{+}(E,0)= B+\dst E\tan \left [ \frac{\pi}{2}E\frac{d}{dE}\right
]\Im mf_{+}(E,0)/E\,
-\dst \frac{2}{\pi
}\sum\limits_{p=0}^{\infty}\frac{C_{+}(p)}{2p+1}\left
(\frac{m_{p}}{E}\right )^{2p}
\end{equation}
with
$$
C_{+}(p)=\frac{e^{-\xi \hat D_{x}}}{2p+1+\hat D_{x}}[ \Im mf_{+}(x,0)-x\Im
mf'_{+}(x,0)]\bigg |_{x=E}
$$
where
$$
f'(x,0)=df/dx,\qquad \hat d_{x}=xd/dx, \qquad \xi=\ln(E/m_{p}).
$$

Similarly the following DDR representation for $f_{-}(E,0)$ is valid

\begin{equation}\label{eq:ddrodd}
\Re ef_{-}(E,0)=\dst \tan \left [ \frac{\pi}{2}E\frac{d}{dE}\right
]\Im mf_{-}(E,0)\,
-\dst \frac{2}{\pi
}\sum\limits_{p=0}^{\infty}\frac{C_{-}(p)}{2p+1}\left
(\frac{m_{p}}{E}\right )^{2p+1}
\end{equation}
where
$$
C_{-}(p)=\frac{e^{-\xi \hat D_{x}}}{2p+\tilde D_{x}} (x/m_{p})\Im mf'_{-}(x,0)\bigg |_{x=E}.
$$
One can prove that $C_{+}(p)$ and $C_{-}(p)$ do not depend on $E$ or $\xi=\ln (E/m_{p})$. Indeed, let us consider an infinitely differentiable function $\Phi(x)$. Then
\begin{equation}\label{eq:Phi}
\begin{array}{lll}
\frac{d}{d\xi}\left (e^{-\xi\hat
d_{x}}\Phi(x)\bigg |_{x=\xi}\right )&=&\frac{d}{d\xi}\left (\sum\limits_{k=0}^{\infty}\frac{(-\xi)^{k}}{k!}\Phi^{(k)}(x)\bigg |_{x=\xi}\right )\\&=&\sum\limits_{k=0}^{\infty}\frac{(-\xi)^{k-1}}{k!}\left
[ -k\Phi^{(k)}(\xi)+(-\xi)\Phi^{(k+1)}(\xi)\right]\\
&=&-\sum\limits_{k=1}^{\infty}\frac{(-\xi)^{k-1}}{(k-1)!}\Phi^{(k)}(\xi)+
\sum\limits_{k=0}^{\infty}\frac{(-\xi)^{k}}{(k)!}\Phi^{(k+1)}(\xi)=0,
\end{array}
\end{equation}
and consequently function $\exp(-\xi \hat d_{x})\Phi (x)|_{x=\xi}$ does not depend on $\xi$.
In our cases $\Phi(\xi)=\frac{1}{2p+1+\hat d_{\xi}}[ \Im mf_{+}(E,0)-E\Im
mf'_{+}(E,0)]$ or $\Phi(\xi)=\frac{1}{2p+\tilde d_{\xi}} (E/m_{p})\Im mf'_{-}(E,0)$. Thus $C_{\pm}(p)$ depend only on $p$ and in spite of their definition in an operator form do not act  on powers of $E/m_{p}$.

On the  one hand the series (\ref{eq:ddreven}) and (\ref{eq:ddrodd}) give the clear expansions in powers of $m_{p}/E$ because $C_{p}$ turn out independent on
energy. Moreover for the most popular Regge type high energy models these constants can be analytically calculated. However on the another hand not only high
energy part of amplitude contributes to $C_{p}$. To have exact values of $C_{p}$ one should know the analytic (differentiable) form of low energy part of
amplitude which is unknown. What we can do is either to add free adjustable parameters instead of $C_{p}$ or to apply uncontrolled
approximation calculating $C_{p}$ from high energy amplitude.

\section{Phenomenology.}
Our aim is to compare the fits of three pomeron models with $\sigma_{t}(s)$ and $\rho(s)$ calculated by three methods: the integral dispersion relation,  the
asymptotic form of the DDR with a subtraction constant and well known  Regge-type parametrization giving simultaneously the both, imaginary and real parts of
scattering amplitude at high energy.

In a light of the soon TOTEM \cite{TOTEM} measurements at LHC we apply  the IDR and DDR methods analyzing here only $pp$ and $\bar pp$ data at $\sqrt{s}\ge 5$ GeV.
However, as we have noted above, we need the low-energy cross sections to take the dispersion integrals.

\subsection{Low-energy part of dispersion integral.}\label{sec:low}
Low-energy total cross sections for $pp$ (151 points) and $\bar pp$ (385 points) interactions at $\sqrt{s}<5$ GeV are explicitly parameterized and
the values of the free parameters are determined from a fit to the data at $s<s_{min}$ (all data on $\sigma$ and $\rho$ are taken from \cite{data}).
We would like to emphasize that aim of such a parametrization is just to have an explicit $E$-dependence of the cross sections in order to perform integration in
Eq. (\ref{eq:dispers II}). Therefore we do not care about  a physical meaning of considered expressions for low energy $\sigma_{tot}(s)$ or number of free
parameters. We just need a model  well describing  the low energy data. In our analysis we take the following low-energy dependencies.

{\bf $\mathbf {pp}$  cross section}
\begin{equation}\label{eq:lowpp}
\sigma_{pp}=\left \{
\begin{array}{llll}
c_{10}+c_{11} p \exp(-(p/c_{12})^{c_{13}}), \qquad & 0&<p\leq &p_{1},\\
c_{20}+c_{21} p + c_{22} p^2+c_{23} p^{3}, \qquad  &p_{1}&\leq p\leq &p_{2},\\
c_{30}+c_{31} p + c_{32} p^2+c_{33} p^{3}+c_{34}p^{4}, \qquad  &p_{2}&\leq p\leq &p_{3},\\
c_{40}+c_{41} p + c_{42} p^2 + c_{43} p^3, \qquad   &p_{3}&\leq p\leq &p_{4},\\
c_{50}+c_{51}( p-p_{4})/p_{4} + c_{52} p^2+c_{53} p^3 , \qquad  &p_{4}&\leq p\leq &p_{min}\\
\end{array}
\right .
\end{equation}

{\bf $\mathbf {\bar pp}$  cross section}
\begin{equation}\label{eq:lowpap}
\sigma^{\bar pp}=\left \{
\begin{array}{llll}
\bar c_{10} + \bar c_{11} p^{\bar c_{12}} \exp (-p/\bar c_{13}),\qquad  & 0& <p\leq &\bar p_{1},\\
\bar c_{20} + \bar c_{21} \exp (-p/\bar c_{22}), \qquad &\bar p_{1}& \leq p\leq & \bar p_{2},\\
\bar c_{30} + \bar c_{31}(p-\bar p_{2})/\bar p_{2}+\bar c_{32} \exp (-p/\bar c_{33})),\qquad  &\bar p_{2}& \leq p\leq & p_{min},\\
\end{array}
\right .
\end{equation}
\noindent
where $p$ is the momentum of proton in the laboratory system, $c_{ik}, \bar c_{ik}$ and $p_{i}, \bar p_{i}$ are adjustable parameters determined from fit to the
data at $\sqrt{s}\leq \sqrt{s_{min}}$ (=5 GeV), $p_{min}$ is the momentum corresponding to $s_{min}$. The value of parameters in Eqs. (\ref{eq:lowpp})  and (\ref{eq:lowpap}) are given in the Table \ref{tab:lowparpp} and  Table \ref{tab:lowparpap}, accordingly.  Not all of these parameters are independent. The equality of cross sections calculated on the left and on the right of each $p_{i}$  and $\bar p_{i}$ ($\lim\limits_{\varepsilon \to
0}\sigma_{pp}(p_{i}-\varepsilon)=\lim\limits_{\varepsilon \to 0}\sigma_{pp}(p_{i}+\varepsilon)$ and similarly for $\sigma^{\bar pp}$) is imposed. These parameters are shown in the Table as fixed.  Let us note that all parameters, besides two ($c_{51}, \bar c_{31}$) which provide the continuity of $\sigma$ at $s_{min}$, are determined independently of the fit at high energy. Therefore they are fixed at all high energy fits.

\FIGURE[h]{\epsfig{file=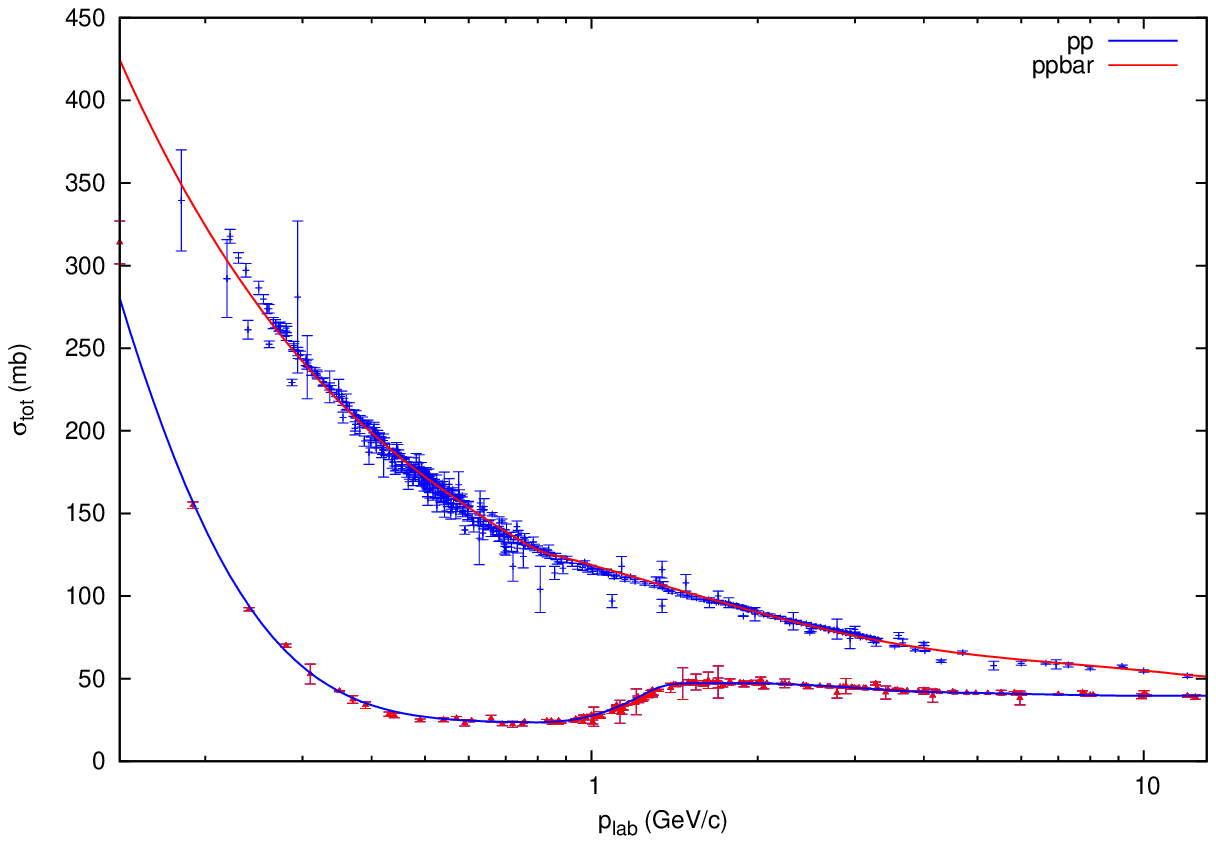, width=0.6 \textwidth }
\caption
{Description of the $pp$ and $\bar pp$ cross sections at low energies, $\sqrt{s}<$ 5 GeV. The data are from the Particle Data Group \cite{data}.}
\label{fig:lowensig} %
}

A quality of   the data description is quite good as it is seen from Fig. \ref{fig:lowensig}. However we would like to comment the obtained $\chi^{2}/dof$ ($\approx$ 5 for $\bar pp$ and $\approx $ 2 for $pp$, ``dof''  is the number of degrees of freedom equaled to number of data points minus number of free parameters). The data are strongly spread around the main group of
points. There are a few points deviating far from the main set of points. Some of them (only 6) individually contribute to $\chi^{2}$ even more than 50. If we
exclude their contribution  to $\chi^{2}$ (not disposing them of fit) we obtain a reasonable value $\chi^{2}/dof \approx 1.5$. In our opinion this quality of the data
description is  acceptable in order to have a sufficiently precise values of the dispersion integral from the threshold up to $\sqrt{s}$=5 GeV.

\TABLE[h]{
\caption{Parameters of the fit (Eqs. \ref{eq:lowpp}) to low energy $pp$-data. (*)Parameter $c_{51}$ is tuned at high-energy fit and takes value in the indicated interval depending on the pomeron model (see Section \ref{sec:high}). }
\label{tab:lowparpp}
\footnotesize{
\begin{tabular}{lrrlrr}
\hline
Parameter&value&error&Parameter&value&error\\
\hline
    $c_{10}$   (mb)                     &       23.35              &      fixed                &     $c_{40}$  (mb)                      &      -3.84     &     fixed     \\
    $c_{11}$   (mb/GeV)            &     140990.0         &     1.3                    &     $c_{41}$  (mb/GeV)             &    80.21    &    4.37    \\
    $c_{12}$   (GeV)                   &    0.015                &    0.006                 &     $c_{42}$  (mb/GeV$^{2}$)  &    -41.28    &    4.77    \\
    $c_{13}$                                 &    0.65                 &       0.10                 &    $c_{43}$    (mb/GeV$^{3}$) &     6.98      &      1.29      \\
    $c_{20}$   (mb)                      &     -19.91             &      fixed                 &     $c_{50}$   (mb)                      &     44.70     &     fixed     \\
    $c_{21}$   (mb/GeV)             &     206.64            &    1.78                   &     $c_{51}$    (mb/GeV)            &     11.83-11.86(*)    &      \\
    $c_{22}$   (mb/GeV$^{2}$)  &     -313.168          &    0.001                 &     $c_{52}$    (mb/GeV$^{2}$) &    0.67    &    0.03    \\
    $c_{23}$   (mb/GeV$^{3}$)  &    153.84             &     3.39                   &     $c_{53}$ (mb/GeV$^{3}$) &     -0.025     &     0.001    \\
    $c_{30}$   (mb)                      &     184.48            &     fixed                  &     $p_{1}$     (GeV)                   &     0.59     &     0.03     \\
    $c_{31}$   (mb/GeV)             &     -326.79           &     3.38                    &     $p_{2}$     (GeV)                   &     1.01     &     0.03     \\
    $c_{32}$   (mb/GeV$^{2}$)  &      41.20             &      1.60                 &     $p_{3}$      (GeV)                  &      1.41      &      0.01\\
    $c_{33}$  (mb/GeV$^{3}$)   &      232.33           &      0.84                 &     $p_{4}$      (GeV)                  &     2.04     &     0.03     \\
    $c_{34}$   (mb/GeV$^{4}$)  &     -103.76           &       0.42              &&&                           \\
     \hline
\end{tabular}
}}

\TABLE[h]{
\caption{Parameters of the fit (Eqs. \ref{eq:lowpap}) to low energy $\bar pp$-data. (*)Parameter $\bar c_{31}$ is tuned at high-energy fit and takes value in the indicated interval depending on the pomeron model (see Section \ref{sec:high}).}
\label{tab:lowparpap}
\footnotesize{
\begin{tabular}{lrrlrr}
\hline
Parameter&value&error&Parameter&value&error\\
\hline
    $\bar c_{10}$ (mb)   &  58.86   &   fixed         &  $\bar c_{30}$ (mb)          &  62.40               & fixed  \\
    $\bar c_{11}$ (mb)   &  66.09  &   8.00          & $\bar c_{31}$ (mb)          &  0.83-0.86(*)    &      \\
    $\bar c_{12}$            &  0.88   &   0.05          &  $\bar c_{32}$ (GeV)        &  100.82            & 0.65\\
    $\bar c_{13}$ (GeV) &  6.74   &   0.12          &  $\bar c_{33}$ (mb)         &  1.59                 & 0.03\\
    $\bar c_{20}$ (mb)   &  91.20  &   fixed          & $\bar p_{1}$  (GeV)       &  0.48               & 0.02 \\
    $\bar c_{21}$ (mb)   &  301.15   &  21.97     & $\bar p_{2}$  (GeV)        &  0.89                & 0.01  \\
    $\bar c_{22}$ (GeV) &   0.38   &   0.04    &  &  & \\
     \hline
\end{tabular}
}}

Thus we perform an overall fit in three steps.
\begin{enumerate}
\item
 The chosen model for high-energy cross-sections is fitted to the data on
the cross sections only at $s>s_{min}$.
 \item The obtained ``high-energy" parameters are fixed. Two ``low-energy" parameters which provide a continuity in the point
$s_{min}$ are determined from the fit at $s<s_{min}$, but with $\sigma_{pp}^{\bar
pp}(s_{min})$  given by the first step.
 \item
There are two possibilities at this step:
\begin{itemize}
\item
The subtraction constant $B_{+}$ is determined from the fit at $s>s_{min}$
with all other parameters kept fixed. Then, without fitting,  the ratios $\rho_{pp}$ and $\rho_{\bar
pp}$ are calculated at all energies above the physical threshold.
\item
Constant $B_{+}$ and high energy parameters simultaneously are determined from the fit at $s>s_{min}$, two parameters from Eqs.~(\ref{eq:lowpp}),
(\ref{eq:lowpap}), $c_{51}$ and $\bar c_{31}$ are tuned to provide a continuity of $\sigma$ at $s=s_{min}$.
\end{itemize}
\end{enumerate}

Investigating the both possibilities at the step 3 we have found that obtained results are in fact coinciding. Therefore in what follows we present the parameter
values and $\chi^{2}$ obtained within the second possibility.

\subsection{High-energy part of dispersion integral. Pomeron models.}\label{sec:high}

We consider three models leading to different asymptotic behavior for the
total cross sections. We start from the explicit parameterization of the
total $pp$ and $\bar pp$ cross-sections, then, to find the ratios of the
real to imaginary parts, we apply the IDR making use of the above described parameterizations (\ref{eq:lowpp}) and (\ref{eq:lowpap})  calculating  the low-energy
part of the dispersion integral.

All the models include the contributions of pomeron, crossing-even and crossing-odd
reggeons (we consider these reggeons as effective ones because it is not
reasonable to take full set of  secondary reggeons, $f, \omega, \rho, a_{2}, ...,$ provided only the $pp$ and
$\bar pp$ data are fitted.)

\begin{equation}\label{eq:sigmod}
Im A_{pp}^{\bar pp}(s,0)= {\cal P}(z)+{\cal R}_{+}(z)\pm {\cal R}_{-}(z),
\end{equation}
where
\begin{equation}\label{eq:reggeon}
{\cal R}_{\pm}(z)=g_{\pm}z^{\alpha_{\pm}(0)}.
\end{equation}
and  $z=|\cos\vartheta _{t}|=2s/(4m_{p}^{2}-t)$ where $\vartheta _{t}$ is the scattering angle in the cross channel. If $t=0$ then $z=s/2m_{p}^{2}=E/m_{p}$.

The real part of amplitude is calculated in two ways. It can be obtained either by IDR (\ref{eq:dispers II}) or by DDR (\ref{eq:ddrev-as}) method. In the both cases
to obtain  total cross section from amplitude we use the optical theorem in the exact form of Eq. (\ref{eq:opt-t-st}).

Besides, we compare our results with the standard Regge-type models. Contribution of Regge pole of the signature $\tau$ ( +1  or -1) to amplitude is
\begin{equation}\label{eq:regge}
A_{R_{\tau}}(s,0)=\eta_{\tau}(\alpha_{R}(0))g_{R}z_{t}^{\alpha_{R}(0)}
\end{equation}
where $\eta_{\tau}(\alpha_{R})$ is a signature factor
\begin{equation}\label{eq:signat}
\dst \eta_{\tau}(\alpha_{R})=\frac{1+\tau \exp(-i\pi \alpha_{R})}{-\sin(\pi \alpha_{R})}
=\left \{
\begin{array}{l}
-\exp(-i\pi \alpha_{R}/2)/\sin(\pi \alpha_{R}/2), \quad \tau=+1,\\
-i\exp(-i\pi \alpha_{R}/2)/\cos(\pi \alpha_{R}/2), \quad \tau=-1.
\end{array}
\right .
\end{equation}
Due to the form  (\ref{eq:signat}) of  signature factor the pomeron and reggeons contributions to $pp, \bar pp$ scattering amplitudes can be written as follows
\begin{equation}\label{eq:-is}
\begin{array}{lll}
A^{\bar pp}_{pp}(s,0)&=&-{\cal P}(-i\tilde s)-{\tilde R}_{+}(-i\tilde s)\mp
i{\tilde R}_{-}(-i\tilde s)
\end{array}
\end{equation}
where $\tilde s=s/s_{0}, s_{0}=1$ GeV$^{2}$ and
\begin{equation}\label{eq:rresidue}
\tilde R_{\pm}(-i\tilde s)=\left \{
\begin{array}{l}
{\cal R}_{+}(-i\tilde s)/\sin(\pi \alpha_{+}/2), \quad \tau=+1,\\
{\cal R}_{-}(-i\tilde s)/\cos(\pi \alpha_{-}/2), \quad \tau=-1.
\end{array}
 \right .
\end{equation}
An advantage of the presentation (\ref{eq:-is}) is that the imaginary parts of the amplitudes (\ref{eq:reggeon}) and (\ref{eq:rresidue}) have the same form. If the asymptotic normalization (\ref{eq:opt-t-as}) is chosen then Eq. \ref{eq:-is} is an ordinary Regge parametrization  in its asymptotic  form.
We denote a fit with such defined amplitudes as ``$-is$ fit".

We present the results of the fit using IDR and the asymptotic form  of  DDR (\ref{eq:ddrev-as}) with
 the standard optical theorem (\ref{eq:opt-t-st})  and compare them with ``$-is$" fit with the asymptotic approximation (\ref{eq:opt-t-as}) of the optical theorem.

Let us now define the pomeron models which we consider and compare.

{\bf Simple pole pomeron model (SP).} In this model, the intercept of the
pomeron is larger than unity. In contrast with well known Donnachie-Landshoff model  \cite{d-l} we add in the amplitude a simple pole (with $\alpha_{P}(0)=1$)
contribution
\begin{equation}\label{eq:SPim}
{\cal P}(E)=g_{0}z+g_{1}z^{\alpha_{\cal P}(0)}.
\end{equation}
The real part of amplitude corresponding to this term and calculated in the DDR method is
\begin{equation}\label{eq:SPre}
Re A_{\cal P}(s,0)=g_{1}\tan(\pi(\alpha_{\cal P}(0)-1))z^{\alpha_{\cal P}(0)}.
\end{equation}

{\bf Dipole pomeron model (DP).} The pomeron in this model is a double
pole in the complex angular momentum plane with intercept $\alpha_{{\cal
P}}(0)=1$.
\begin{equation}\label{eq:DPim}
{\cal P}(E)=g_{0}z+g_{1}z \ln z.
\end{equation}
In the DDR method
\begin{equation}\label{eq:DPre}
Re A_{\cal P}(s,0)=g_{1}z\pi/2.
\end{equation}

{\bf Tripole pomeron model (TP)}. This pomeron is the hardest complex
$j$-plane singularity allowed by unitarity, it is a triple pole at $t=0$ and
$j=1$.
\begin{equation}\label{eq:TPim}
{\cal P}(E)=g_{0}z+g_{1}z\ln z+g_{2}z\ln^{2}z.
\end{equation}
In the DDR method
\begin{equation}\label{eq:TPre}
Re A_{\cal P}(s,0)=g_{1}z\pi /2+g_{2}z\pi \ln z.
\end{equation}

In all considered above pomeron models real part of amplitude calculated by DDR is
\begin{equation}\label{eq:realDDR}
Re A_{pp}^{\bar pp}(s,0)=B/(2m_{p})+Re A_{\cal P}(s,0)\pm ReA_{\cal R_{\pm}}(s,0)
\end{equation}
where $Re A_{\cal P}(s,0)$ is given by Eqs. (\ref{eq:SPre}), (\ref{eq:DPre}), (\ref{eq:TPre}) and
\begin{equation}\label{eq:reggre}
ReA_{\cal R_{\pm}}(s,0)=\left \{
\begin{array}{l}
-z\cot (\pi \alpha_{R_{+}}(0))z^{\alpha_{R_{+}}(0)},\\
z\tan (\pi \alpha_{R_{-}}(0))z^{\alpha_{R_{-}}(0)}.
\end{array}
\right.
\end{equation}

\section{Fit results, discussion and conclusion.}

The values of $\chi^{2}$ and adjustable parameters  obtained in fits are presented in the Tables \ref{tab:chi2}, \ref{tab:SPpar}, \ref{tab:DPpar} and \ref{tab:TPpar}. There are 238 experimental points in the PDG data set for $pp$ and $\bar pp$ interactions \cite{data} which were used in the fits. It contains 104 (59) points for $pp$ ($\bar pp$) cross section and 64 (11) points for $pp$ ($\bar pp$) ratio $\rho$.
The curves obtained  for the total cross sections in the three considered pomeron models, if IDR method is applied,  are shown in the Fig.~\ref{fig:allensig}. To compare the methods IDR, DDR and $-is$ we have plotted the $\sigma_{tot}$ and $\rho$ curves in each considered pomeron models in Figs.~\ref{fig:SP-st-3met} - \ref{fig:TP-rho-3met}.

\TABLE[h]{
\caption{\label{tab:chi2}The values of the $\chi^{2}$ per degree of freedom ($\chi^{2}/dof$) obtained in the various pomeron models and through the
different methods for the calculation of the ratio~$\rho$. The both $\sigma_{tot}$ and $\rho$ are taken into account.}
\small{
\begin{tabular}{lccc}
\hline
Pomeron model  & IDR & DDR &  $"-is"$ \\
  \hline
 Simple pole & 1.096 & 1.102 & 1.135 \\
 Double pole & 1.103 & 1.108 & 1.131 \\
 Triple pole & 1.096 & 1.113 & 1.135  \\
\hline
\end{tabular}}
}

Let us draw attention to some details of the fit and compare specific features of the considered pomeron models and methods for $\rho$ calculations.

All considered pomeron models give a good description of the
data on $\sigma_{tot}$ and $\rho$ as one can see from the Table \ref{tab:chi2}. However, specifically  the fit complemented  by IDR method  is preferable within
the all models. Moreover the Figs.~\ref{fig:SP-rho-3met}, \ref{fig:DP-rho-3met}, \ref{fig:TP-rho-3met} show that $\rho$ calculated by IDR method is in a visibly better agreement  with the data at low energy which are outside of fit range $\sqrt{s}>$ 5 GeV.
We would like to note that the values of $\rho$ calculated using DDR are deviated from those calculated with IDR even at $\sqrt{s} \lesssim 7 -8$
GeV , notably, in order to have more correct values of the $\rho$ at such energies, one must use the IDR rather than to calculate $\rho$ by the DDR or ``$-is$'' methods.

\FIGURE[h]{
\includegraphics[scale=0.8]{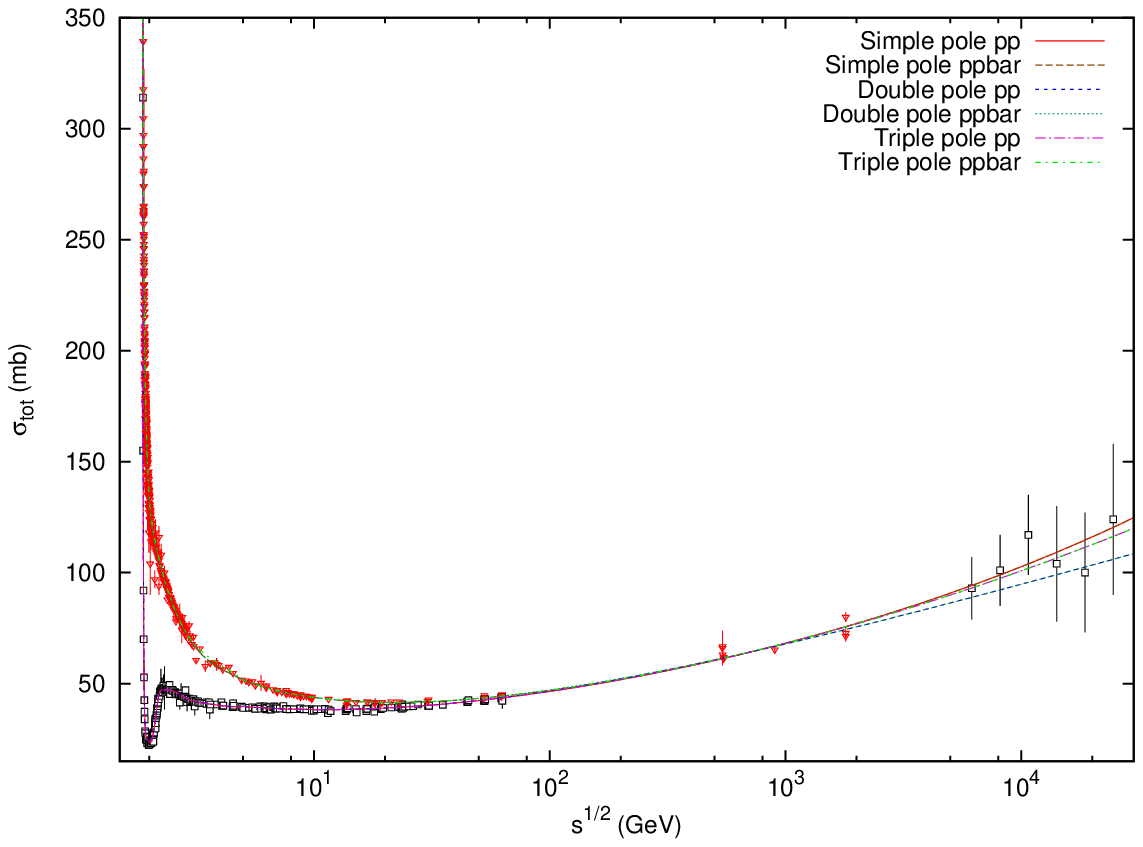}
\caption{\label{fig:allensig}Description of the $pp$ and $\bar pp$ cross sections at low energies and at  $\sqrt{s}>$ 5 GeV in three pomeron models. Explanation of the lines is given in the figure. The data are from the Particle Data Group \cite{data}.}}

\DOUBLEFIGURE[htb]
{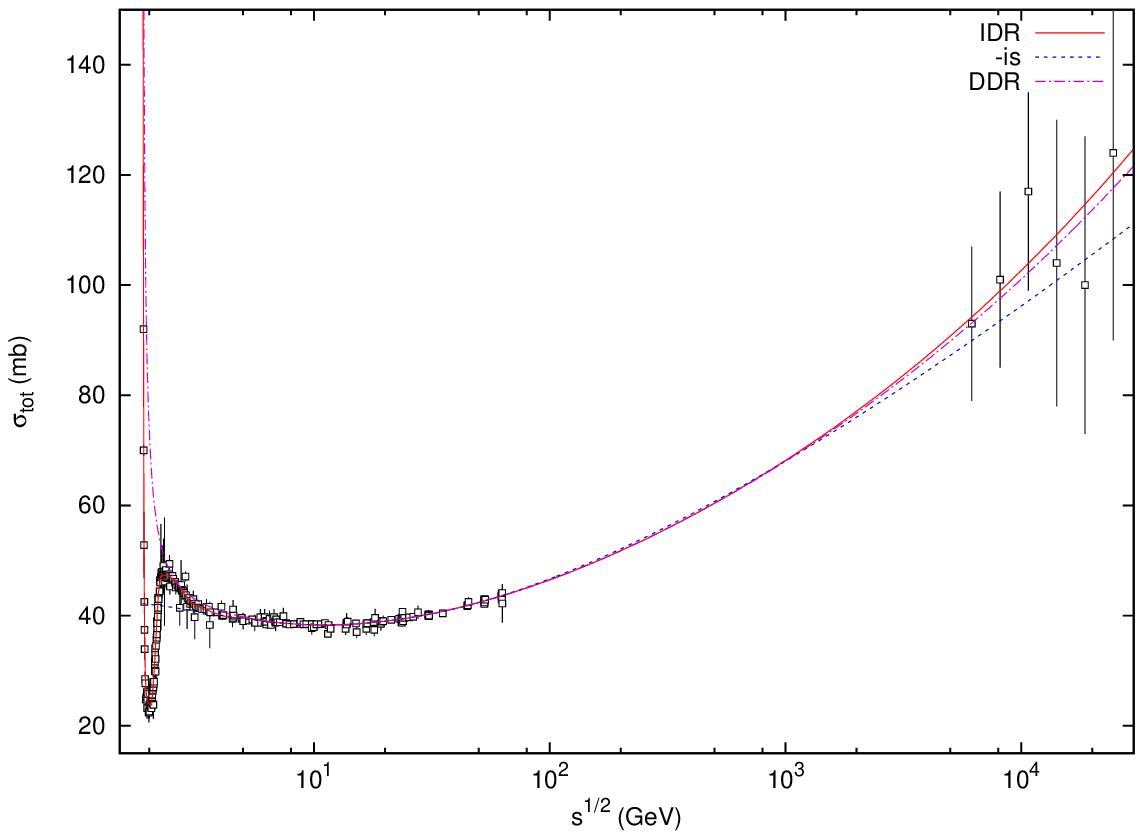, width=.5\textwidth}
{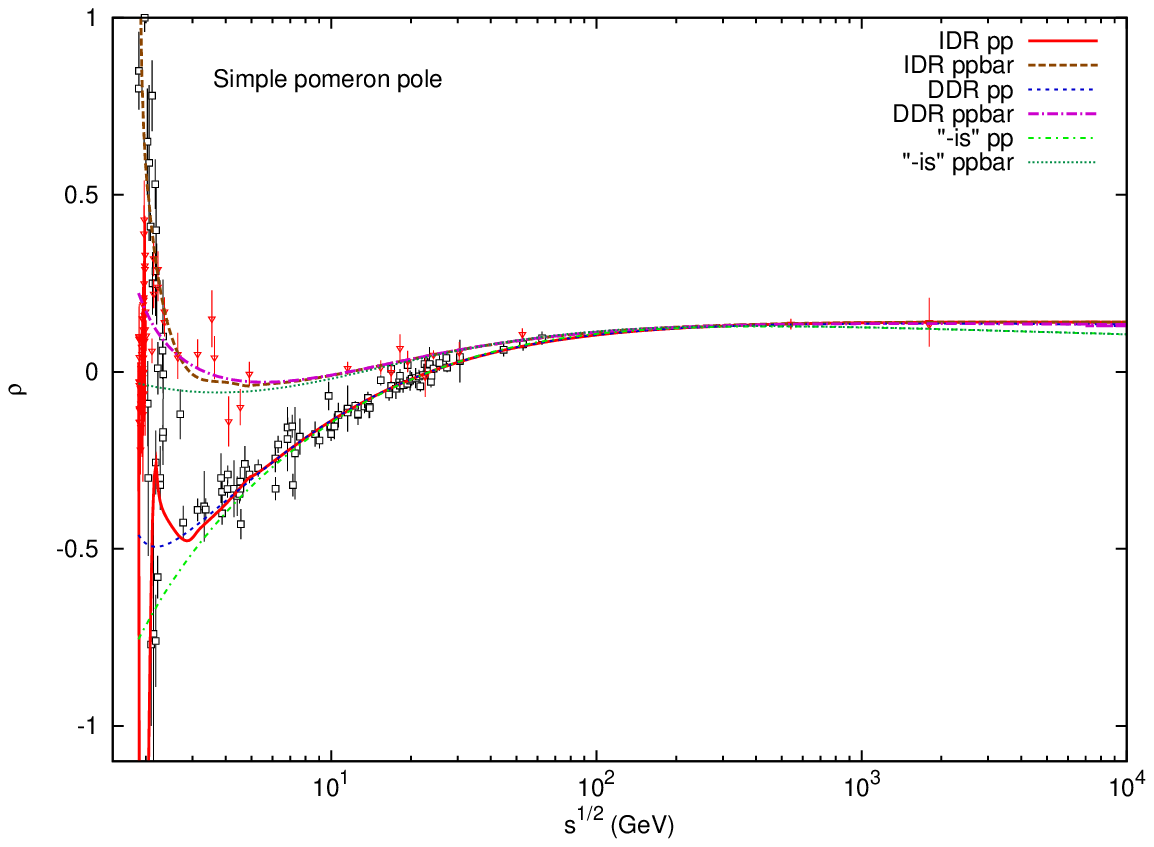, width=.5\textwidth}
{\label{fig:SP-st-3met} Description of $\sigma_{tot}$ in the simple pole pomeron models by three methods: IDR, DDR, ``$-is$''. The data are from the Particle Data Group \cite{data}.}
{\label{fig:SP-rho-3met} Description of $\rho$ in the simple pomeron models by three methods: IDR, DDR, ``$-is$''. The data are from the Particle Data Group \cite{data}.}

\DOUBLEFIGURE[htb]
{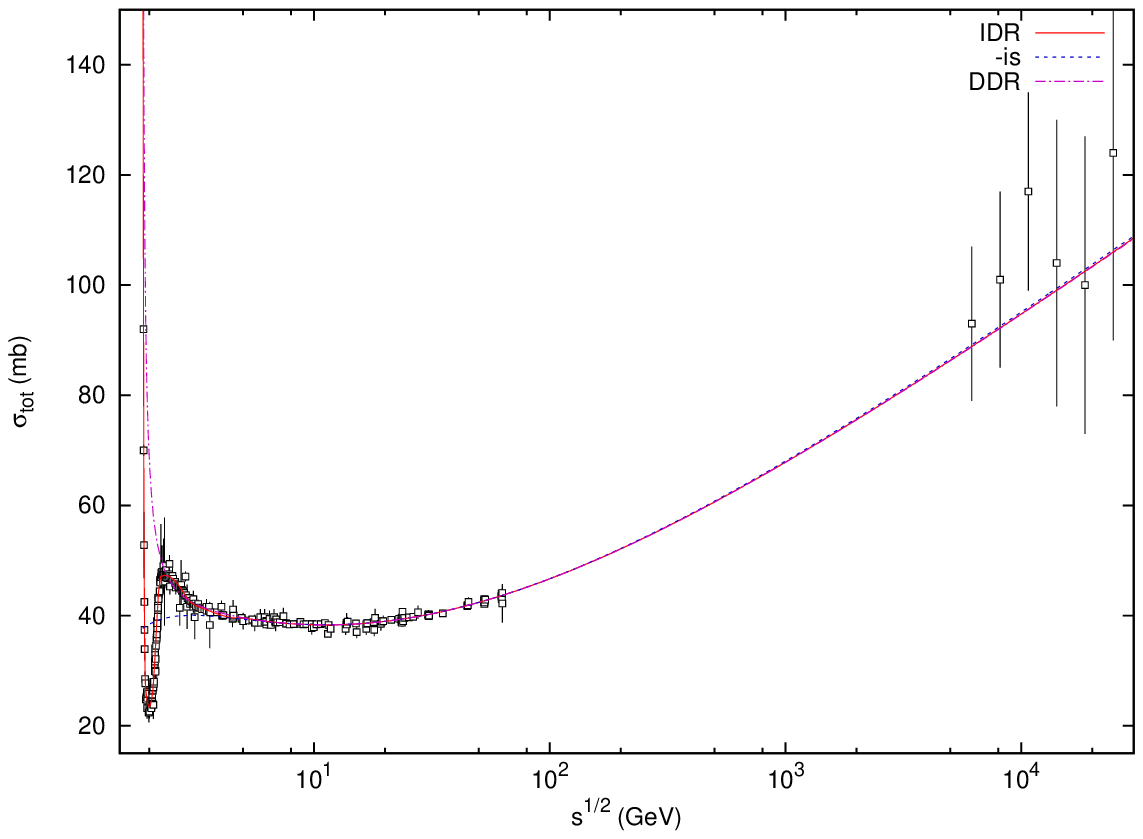, width=.5\textwidth}
{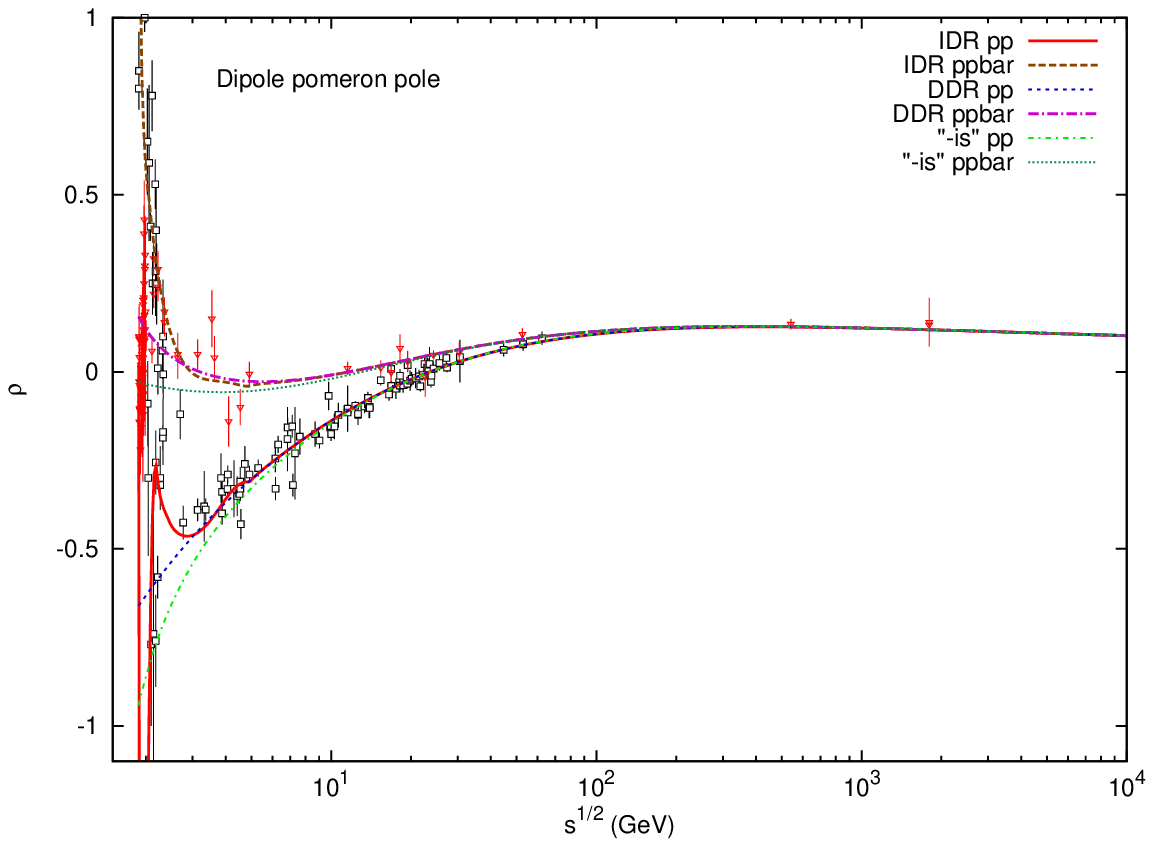, width=.5\textwidth}
{\label{fig:DP-st-3met} The same as in Fig.~\protect \ref{fig:SP-st-3met}  but for the double pole pomeron model.}
{\label{fig:DP-rho-3met} The same as in Fig.~\protect \ref{fig:SP-rho-3met} but for the double pole pomeron model.}

\DOUBLEFIGURE[htb]
{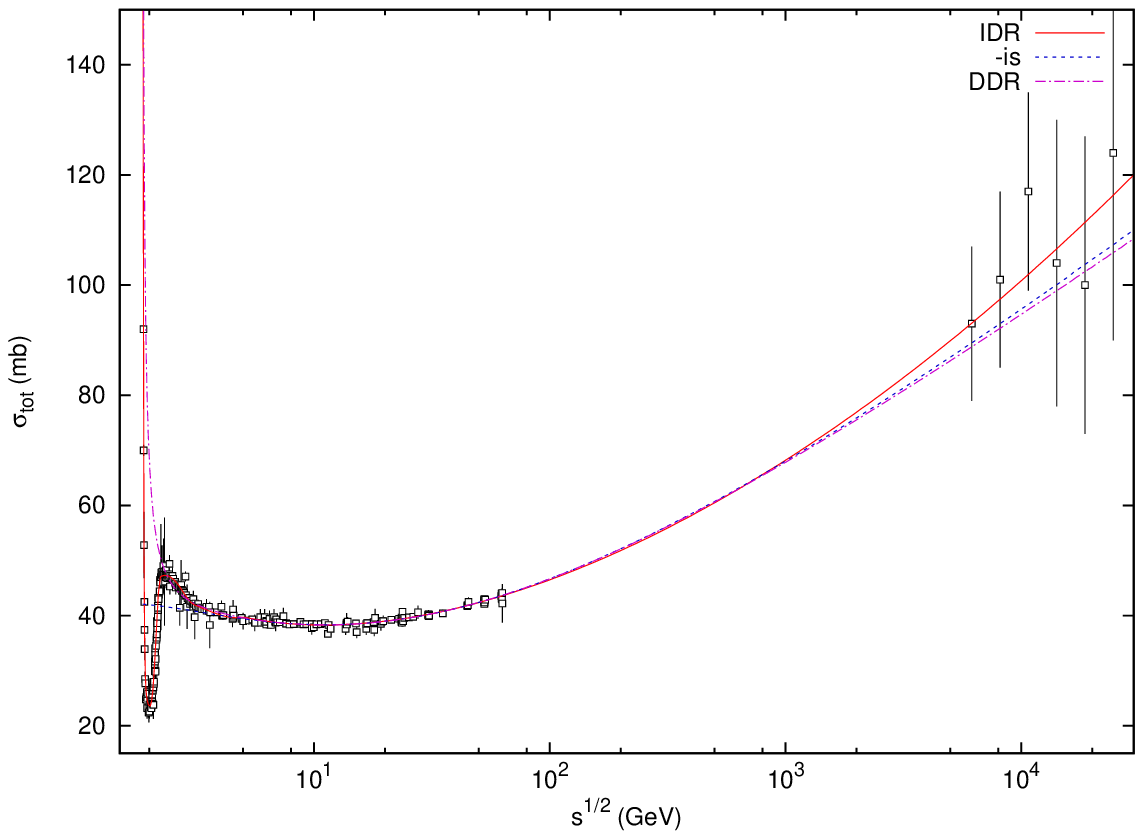, width=.5\textwidth}
{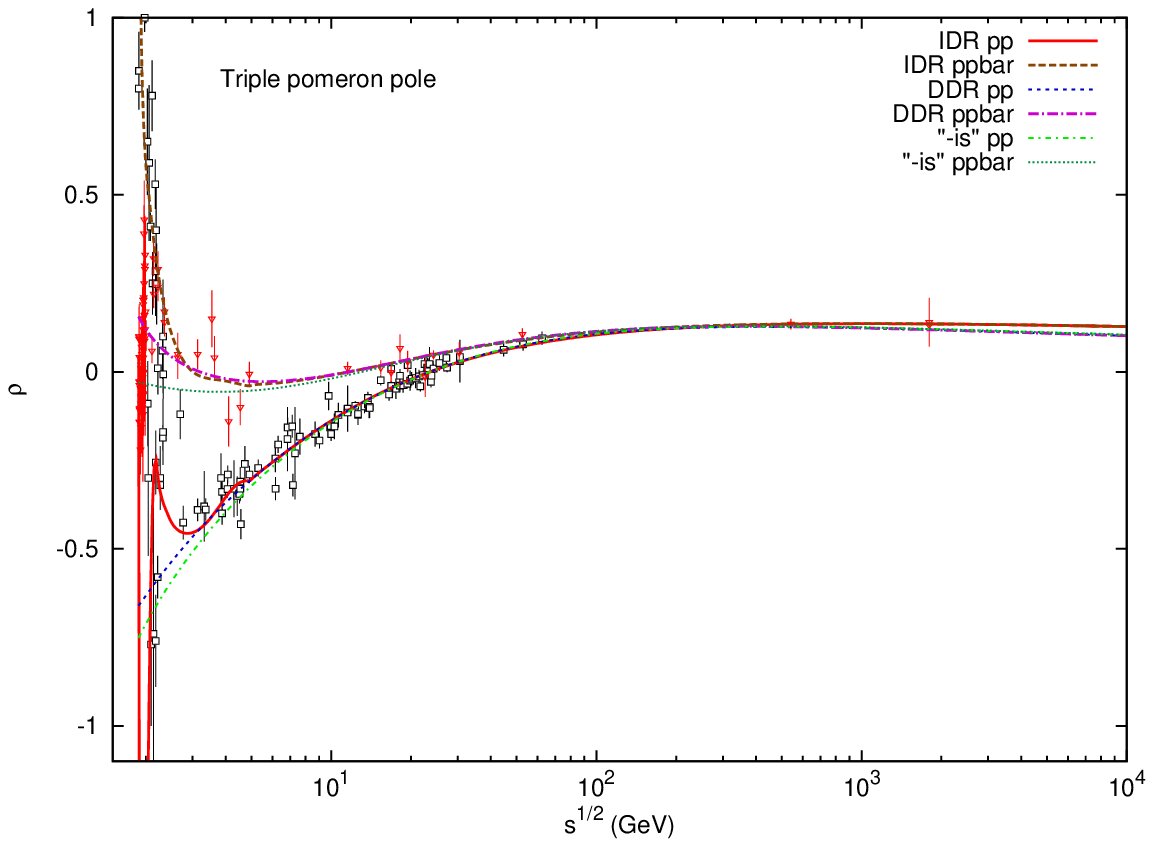, width=.5\textwidth}
{\label{fig:TP-st-3met} The same as in  Fig.~\protect \ref{fig:SP-st-3met} but for the triple pole pomeron model.}
{\label{fig:TP-rho-3met} The same as in Fig.~\protect \ref{fig:SP-rho-3met} but for the triple pole pomeron model.}

\DOUBLEFIGURE[htb]
{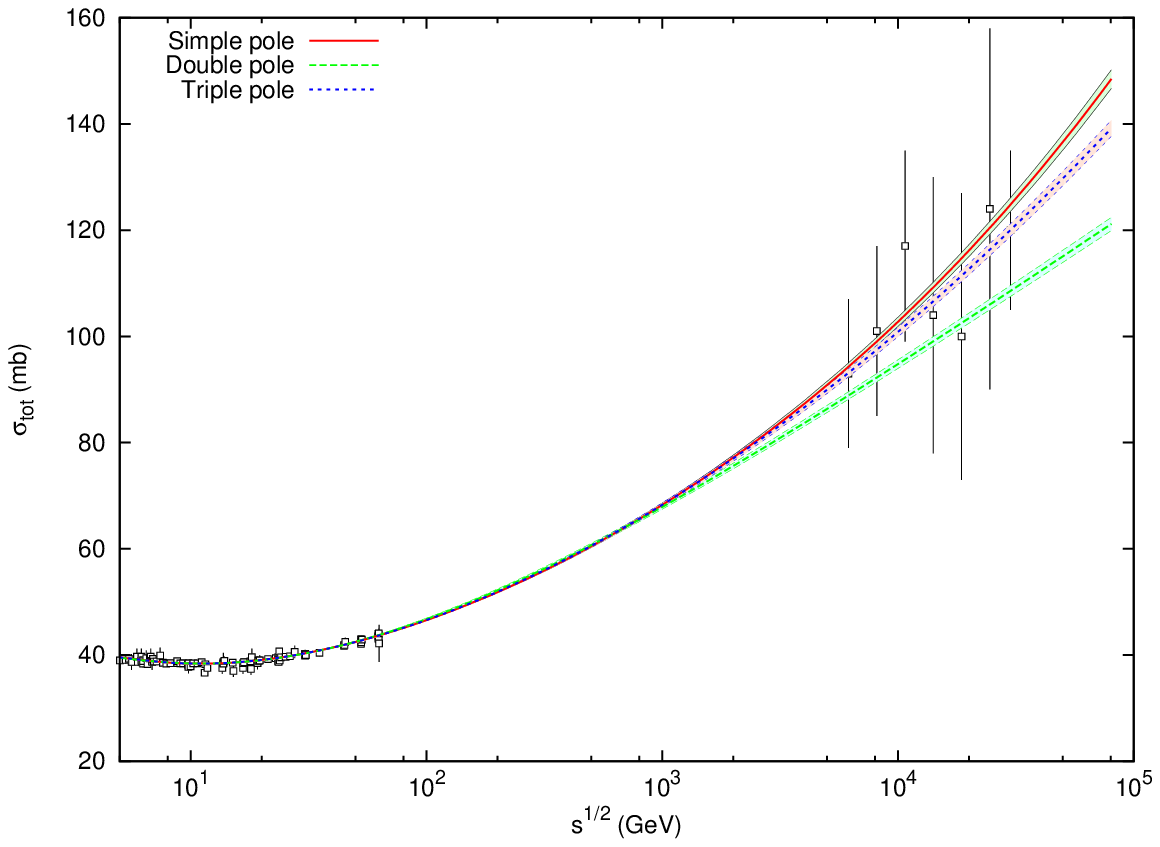, width=.5\textwidth}
{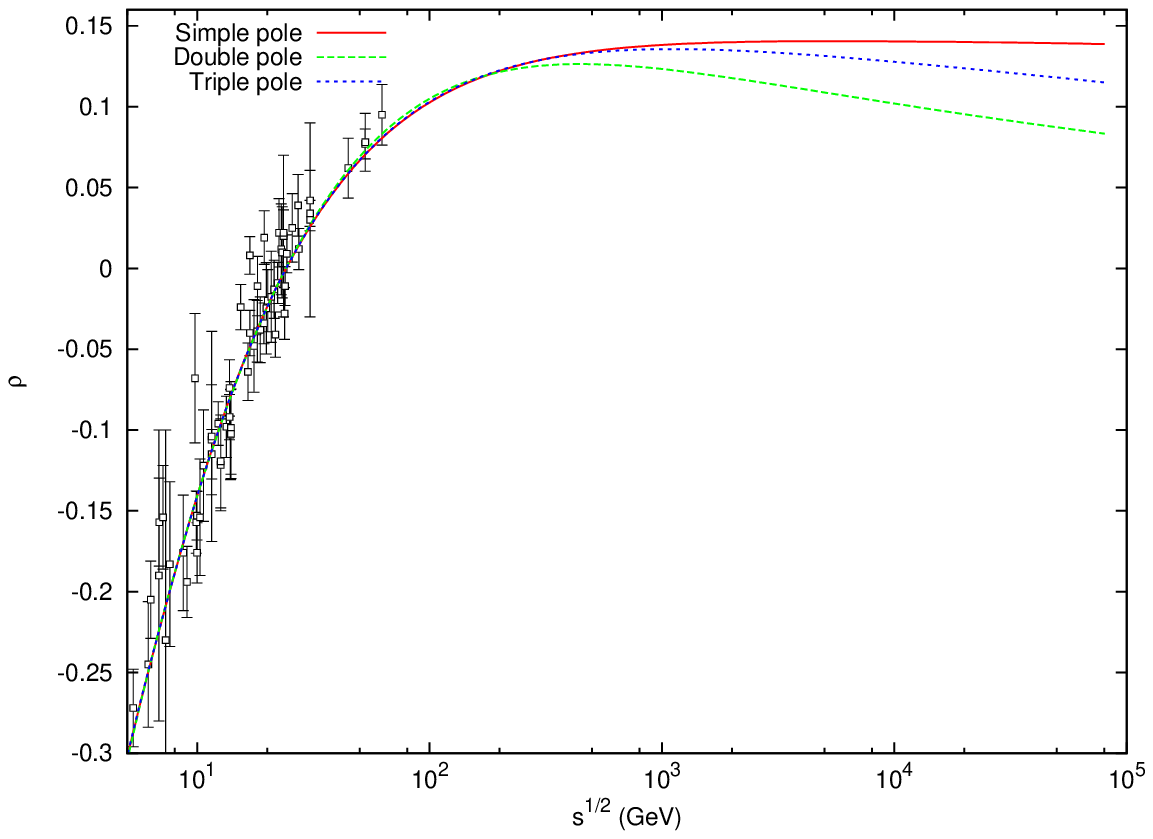, width=.5\textwidth}
{ \label{fig:highen3mod}Description of $pp$  cross sections  at $\sqrt{s}>$ 5 GeV in three pomeron models ($\rho$ is calculated by the IDR method). The bands for $\sigma_{tot}$ are shown.}
{ \label{fig:aaa}The ratios $\rho$  calculated by the IDR method in three pomeron models.}

\TABLE[h]{
{\footnotesize
\caption{Parameters of the simple pomeron model in three methods of $\rho$ calculations.}
\label{tab:SPpar}
\begin{tabular}{lccc}
\hline
  Parameters& IDR & DDR & ''$-is$``  \\
\hline
  $\alpha_{\cal P}(0)$&   1.086 $\pm$   0.001    & 1.075 $\pm$  0.001       &      1.013 $\pm$ 0.0001 \\
 $g_{0}$                   &  -20.84 $\pm$   1.11         &  -62.02 $\pm$   1.13      &  -1071.1 $\pm$ 0.7      \\
 $g_{1}$                   & 105.30 $\pm$   0.74         &   135.45 $\pm$   0.78    &     1026.8 $\pm$ 0.6       \\
  $\alpha_{+}(0)$     &       0.652 $\pm$   0.005   &  0.669 $\pm$   0.004     &     0.776 $\pm$ 0.001  \\
$g_{+}$                    &   235.12 $\pm$   2.69       &  243.26 $\pm$   2.55     &     253.29 $\pm$ 1.40   \\
$\alpha_{-}(0)$        &     0.4630 $\pm$   0.01       &  0.464 $\pm$   0.010     &     0.450 $\pm$ 0.010   \\
 $g_{-}$                    &  107.11 $\pm$   4.35        &  106.92 $\pm$   4.34     &   115.96 $\pm$ 3.88  \\
 $B (GeV)$              &    -35.76 $\pm$   7.11       &   228.66 $\pm$  66.97   &  \\
\hline
$\chi^{2}/dof$ &1.096 & 1.102 & 1.135\\
\hline
\end{tabular}
}}

\TABLE[h]{
\caption{Parameters of the double pomeron model in three methods of $\rho$ calculations.}
\label{tab:DPpar}
{\footnotesize
\begin{tabular}{lccc}
\hline
  Parameters& IDR & DDR & ''$-is$``  \\
\hline
 $g_{0}$                   &  -121.54 $\pm$   3.10         &  -120.193 $\pm$ 29.53           & -78.34 $\pm$ 15.42   \\
 $g_{1}$                   & 30.18 $\pm$   1.96        &    30.09 $\pm$  1.88                &  17.17 $\pm$  0.96     \\
  $\alpha_{+}(0)$     &       0.797 $\pm$   0.02   &     0.796 $\pm$  0.015             &     0.794 $\pm$  0.013\\
$g_{+}$                    &   420.31 $\pm$   2.70       &    419.13 $\pm$  25.75           & 270.12 $\pm$  12.7   \\
$\alpha_{-}(0)$        &     0.465 $\pm$   0.012       &      0.465 $\pm$  0.012            &   0.451 $\pm$  0.012   \\
 $g_{-}$                    &  106.25 $\pm$   5.15        &   106.21 $\pm$  5.18             & 87.91 $\pm$  4.92  \\
 $B (GeV)$              &    -35.99 $\pm$   7.28       &     111.62 $\pm$  72.96            &  \\
\hline
$\chi^{2}/dof$ &1.103 & 1.108 & 1.131\\
\hline
\end{tabular}
}}

\TABLE[h]{
\caption{Parameters of the triple pomeron model in three methods of $\rho$ calculations.}
\label{tab:TPpar}
{\footnotesize
\begin{tabular}{lccc}
\hline
  Parameters& IDR & DDR & ''$-is$``  \\
\hline
 $g_{0}$                   &  139.90 $\pm$   1.11         &  -120.17 $\pm$   29.03      &  -75.41 $\pm$ 0.66      \\
 $g_{1}$                   & -1.77 $\pm$   0.21         &   30.09 $\pm$   1.84    &     16.87 $\pm$ 0.11       \\
 $g_{2}$                   & 1.09 $\pm$   0.02         &   $1\cdot 10^{-8} \pm 3\cdot 10^{-3}$     &     0.009 $\pm$ 0.01        \\
  $\alpha_{+}(0)$     &       0.589 $\pm$   0.008   &  0.796 $\pm$   0.014     &     0.793 $\pm$ 0.001  \\
$g_{+}$                    &   183.61 $\pm$   3.61       &  419.11 $\pm$   25.32     &     267.27 $\pm$ 1.20   \\
$\alpha_{-}(0)$        &     0.463 $\pm$   0.010       &  0.465 $\pm$   0.012     &     0.451 $\pm$ 0.009   \\
 $g_{-}$                    &  107.25 $\pm$   4.42        &  106.22 $\pm$   5.18     &   87.92 $\pm$ 3.85  \\
 $B (GeV)$              &    -33.17 $\pm$   7.12      &   111.64 $\pm$  72.79   &  \\
\hline
$\chi^{2}/dof$ &1.096 & 1.113 & 1.135\\
\hline
\end{tabular}
}}

While the data on $\sigma$ are described with $\chi^{2}/N_{p} \approx 0.85 - 0.95$, the data on $\rho$ are described less well, with a
$\chi^{2}/N_{p}\approx 1.6$ in $pp$ case (0.4 - 0.45 for $\bar pp$). We think it occurs because of insufficient  quality of the $\rho$ data.

Neglecting the subtraction constant and using the asymptotic normalization (\ref{eq:opt-t-as}) in a non-asymptotic domain, may cased  the
conclusion of \cite{COMPETE} concerning the simple pole model. It was excluded from the list of the best models. Including  in (\ref{eq:SPim}) the sub-asymptotic term  considerably improves the description of $\rho$ as well as $\sigma_{tot}$ at energy slightly higher of $\sqrt{s}=5$ GeV and renew the status of the model as one of the best phenomenological models in spite of its contradiction  with unitarity.
Let us note the important differences between COMPETE's and our approaches. Firstly, in our pomeron models all amplitudes are written (when IDR or DDR is applied) as functions of Regge variable  $|z_{t}|=E/m_{p}$ with the exact form (Eq.~\ref{eq:opt-t-st}) of optical theorem while in the COMPETE fits $\sigma$ and $\rho$ are  functions of $s$ (in fact, to derive expressions for $\rho$ asymptotic form (Eq.~\ref{eq:opt-t-st}) was used). Secondly COMPETE made the global fit including the $\pi^{\pm}p, K^{\pm}p, \gamma p, \gamma \gamma$ and $\Sigma^{-}p$ data. We have considered here only $pp$ $\bar pp$ data. We intend to extend our IDR and DDR analysis for all mentioned data. Then it would be possible to compare these two approaches.

We would also like to draw attention to the parameter values obtained in the simple and triple pole pomeron models by the ``$-is$'' fit (see the Tables
\ref{tab:SPpar}, \ref{tab:DPpar} and \ref{tab:TPpar}) as well as in the triple model with the  DDR method. It is interesting that in fact triple pomeron reproduces the
dipole pomeron because $g_{2}\approx 0$ while the simple pole with small value of $\alpha_{\cal P}(0)-1$ is quite closed to the dipole pomeron.

Total cross sections and ratio $\rho$ in the considered models  behave almost identically at energy 5 GeV $<\sqrt{s}\lesssim 1$ TeV, but at higher energy they
 deviate from each other (see Fig.~\ref{fig:highen3mod}).  Predictions of the $pp$ total cross section (with the errors) and of the ratio $\rho_{pp}$ in region of
LHC energies are shown in the Figs.~\ref{fig:lhc-s3mod}, \ref{fig:lhc-r3mod} for three considered pomeron models.  Numerical values for $\sigma_{pp}$ and $\rho_{pp}$ at 7 TeV and
14 TeV are given in the Table \ref{tab:lhc-predict}.

\DOUBLEFIGURE[htb]
{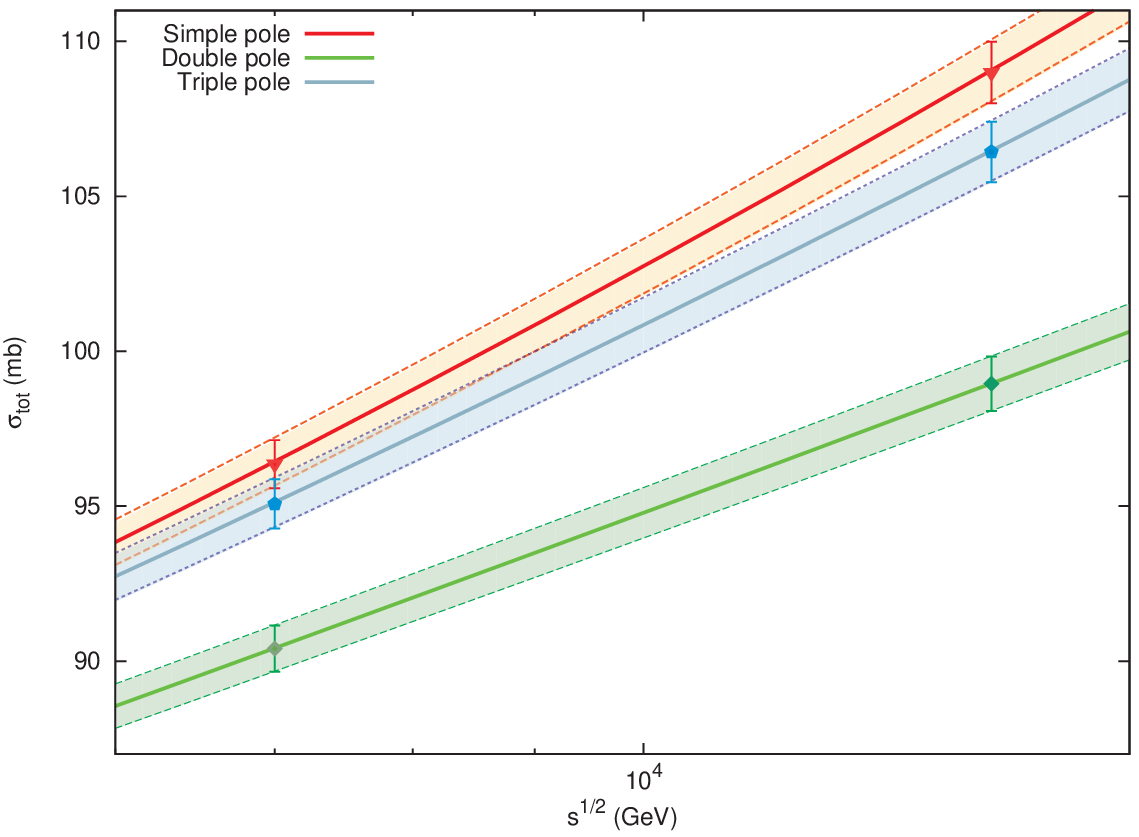, width=.48\textwidth}
{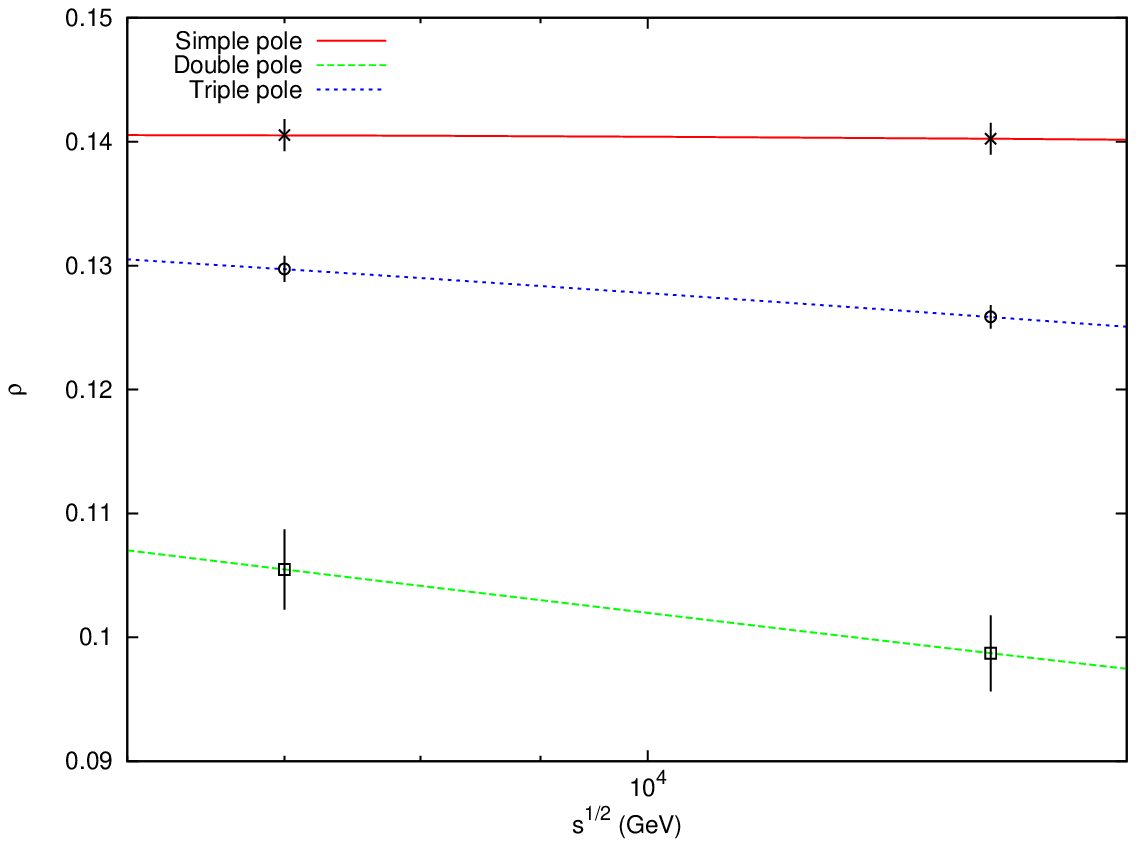, width=.52\textwidth}
{ \label{fig:lhc-s3mod}Predictions of $pp$  cross sections  at $\sqrt{s}=$ 7 TeV and 14 TeV in three pomeron models.}
{ \label{fig:lhc-r3mod}Predictions of  $\rho$ at $\sqrt{s}=$ 7 TeV and 14 TeV in three pomeron models.}

\TABLE[h]{
\caption{Predicted values of $\sigma_{pp}$ and $\rho_{pp}$ at LHC energies in three pomeron models}
\label{tab:lhc-predict}
\begin{tabular}{ccccc}
\hline
&&Simple pole  & Double pole & Triple pole \\
\hline
 $\sqrt{s}$=7  TeV &
 $\sigma_{pp}$ & 96.36$\pm $0.77 & 90.40$\pm $0.75 & 95.07$\pm $0.79 \\
& $\rho_{pp}$ & 0.141$\pm $0.001 & 0.106$\pm $0.003 & 0.130$\pm $0.001\\
\hline
 $\sqrt{s}$=14  TeV &
 $\sigma_{pp}$  & 108.99$\pm $0.99 &98.96$\pm $0.88  & 106.43$\pm $0.98 \\
& $\rho_{pp}$  & 0.140$\pm $0.001 & 0.099$\pm $0.003 & 0.126$\pm $0.001 \\
\hline
\end{tabular}
}

Summarizing we would like to emphasize that
\begin{itemize}
\item
the integral dispersion relations  for  $pp$ and $\bar pp$ forward elastic scattering amplitudes expressing their analyticity allow not only to analyze high energy
behaviour of $\sigma_{pp}^{\bar pp}(s)$ and $\rho_{pp}^{\bar pp}(s)$ but also lead to a  more proper description of the data on  $\rho_{pp}^{\bar pp}(s)$ at low
energy;
\item
the integral and derivative dispersion relations together with exact form of the optical theorem give a better agreement with the data at $\sqrt{s}\lesssim 7 - 8$
GeV;
\item
predicted  $pp$ total cross sections at LHC energies deriving by IDR method are in the interval 90 - 97 mb at $\sqrt{s}$=7 TeV and in the interval 98 - 109 mb
at $\sqrt{s}$=14 TeV;
\item
predicted values for $\rho$ are located in the intervals 0.103 - 0.142 at $\sqrt{s}$=7 TeV and 0.096 - 0.141 at $\sqrt{s}$=14 TeV;
\item
if the TOTEM experiment \cite{TOTEM} achieves the declared precision 1\% for $\sigma_{pp}$ it would be possible to discriminate the considered models and therefore to chose from them the most adequate one.
\end{itemize}

Authors thank J.R. Cudell for fruitful discussions and critical remarks.

\end{document}